\documentclass[showpacs,pre,twocolumn]{revtex4}%
\usepackage{amsfonts}
\usepackage{amsmath}
\usepackage{amssymb}
\usepackage{graphicx}%
\begin{document}
\preprint{ }
\title[Short title for running header]{Derivation of the residence time for kinetic Monte Carlo simulations}
\author{Clinton DeW. Van Siclen}
\email{clinton.vansiclen@inl.gov}
\affiliation{Idaho National Laboratory, Idaho Falls, Idaho 83415, USA}

\begin{abstract}
The kinetic Monte Carlo method is a standard approach for simulating physical
systems whose dynamics are stochastic or that evolve in a probabilistic
manner. Here we show how to calculate the system time for such simulations.

\end{abstract}
\volumeyear{year}
\volumenumber{number}
\issuenumber{number}
\eid{identifier}
\date{November 16, 2007}
\startpage{1}
\endpage{102}
\maketitle

\section{Introduction}

In a kinetic Monte Carlo (kMC) simulation, a model system is moved from state
to state according to probabilistic rules. A nice example is adatom diffusion
on a surface: in this case the \textquotedblleft system\textquotedblright\ is
the adatom plus surface, and a \textquotedblleft state\textquotedblright\ has
the adatom at a particular surface site. The probabilistic rules for moving
the adatom from one site to another (i.e., the jump rates) are obtained from
molecular dynamics (MD) calculations, which give the energy barriers over
which the moves occur. Subsequently, a separate kMC simulation moves the
adatom from site to site over the surface, and thereby enables calculation of
the adatom diffusion coefficient, $D=\left\langle x^{2}\right\rangle /(4t)$,
from distance $x$ traveled by the adatom over the time interval $t$.

This example is typical in that the kMC simulation gives a macro- or
mesoscopic quantity of interest (in this case the adatom diffusion
coefficient), using information obtained from the microscopic or atomic scale.
In fact, molecular dynamics calculations alone are impractical for simulating
adatom diffusion over a surface: the adatom at room temperature attempts a
jump\ every $10^{-13}$ seconds, while it is successful once every $10^{-5}$
seconds (so in MD simulations it is the numerous unsuccessful attempts at many
sites, rather than the rare successful one, that enables the energy landscape
to be mapped and jump rates determined).

The great virtue of the kMC method is that it enables simulation of physical
processes that involve very disparate time scales. It is then essential that
the actual time in the physical system be correctly reproduced by the
\textit{residence times} in the states of the model system, or equivalently,
by the time between events in the model system.

While equivalent statements, the different semantics suggests the ability of
the kMC method to obtain properties of an equilibrium system \textit{as well
as} to model the dynamic evolution of a non-equilibrium system. In the case of
equilibrium systems, each state may be visited many times, so that the
residence times in the states are proportional to the equilibrium populations
of the physical states (for example, the adatom on a finite surface may be
found at sites in proportion to the residence times at the corresponding
states in the model system). More generally the kMC method is used to model
the evolution of a non-equilibrium system (that is, where a state once left
can never be returned to), so this case will motivate the derivation of the
residence time presented below. However, the mathematics apply to equilibrium
systems as well, as will be made apparent at the end of the derivation.

\section{Derivation of the residence time}

Consider a collection $\{j\}$ of elements, where each element $j$ has an
expected lifetime\ $\tau_{j}$. The kMC method follows from the assumption that
the values $\{\tau_{j}\}$ are time-independent (i.e., that the transition
rates $\{\tau_{j}^{-1}\}$ from the current state to accessible states are
time-independent). Then the probability $p_{j}(t)dt$ that the particular
element $j$ will fail during the infinitesimal time interval $[t,t+dt]$ is
given by the probability that element $j$ will \textit{not} fail prior to time
$t$, multiplied by the probability $\tau_{j}^{-1}dt$ that it \textit{will}
fail during the subsequent time interval $dt$; that is,%
\begin{equation}
p_{j}(t)dt=\left[  1-\int\limits_{0}^{t}p_{j}(t^{\prime})dt^{\prime}\right]
\frac{dt}{\tau_{j}}\text{.} \label{e1}%
\end{equation}
This equation simplifies to%
\begin{equation}
p_{j}(t)=\left[  1-\int\limits_{0}^{t}p_{j}(t^{\prime})dt^{\prime}\right]
\frac{1}{\tau_{j}}\text{.} \label{e2}%
\end{equation}
Taking the derivative of each side of Eq. (\ref{e2}) with respect to $t$ and
integrating produces the exponential probability density function (PDF)%
\begin{equation}
p_{j}(t)=\frac{1}{\tau_{j}}\exp\left[  -\frac{t}{\tau_{j}}\right]  \text{.}
\label{e3}%
\end{equation}
Specifically, $p_{j}(t)$ is the \textit{distribution} of failure times for
element $j$; the \textit{average} value of the failure time is $\left\langle
t_{j}\right\rangle =%
{\textstyle\int\nolimits_{0}^{\infty}}
tp_{j}(t)dt=\tau_{j}$.

To perform a simulation, we need to randomly select a time $t$ from this
distribution. The formula for converting a random number $x$ taken from the
\textit{uniform} probability distribution $P(x)=1$ (such $x$ values are
produced by the standard random number generators) to the corresponding $t$
value is derived as follows. The probabilities $p(t)dt$ and $P(x)dx$ must be
equal, so $p(t)dt=dx$. Thus%
\begin{equation}
x(t)=\int\limits_{0}^{t}p(t^{\prime})dt^{\prime}=1-\exp\left[  -\frac{t}%
{\tau_{j}}\right]  \text{.} \label{e4}%
\end{equation}
Inverting this expression then gives the desired relation between $x$ and $t$,%
\begin{equation}
t=\tau_{j}[-\ln(1-x)] \label{e5}%
\end{equation}
with $x$ randomly chosen from the interval $[0,1)$.

In this way we get the set $\{t_{j}\}$ of element failure times. If the
elements $\{j\}$ are completely independent of one another, they fail
according to the time ordering of the set $\{t_{j}\}$.

If, however, the elements are not independent, the remaining values
$\{\tau_{j}\}$ are altered with each failure. Consider that the smallest
member of the set $\{t_{j}\}$ is $T_{1}$, so the first element to fail does so
at time $T_{1}$. Which is the next element to fail, and when does that occur?
If element $j$ is still functioning at time $T_{1}$, the PDF%
\begin{equation}
p_{j}(t-T_{1})=\frac{1}{\tau_{j}^{\prime}}\exp\left[  -\frac{(t-T_{1})}%
{\tau_{j}^{\prime}}\right]  \label{e6}%
\end{equation}
where $\tau_{j}^{\prime}$ is the new lifetime of element $j$ for time
$t>T_{1}$. Equation (\ref{e6}) gives the distribution of times $t>T_{1}$ at
which element $j$ fails, or equivalently, the distribution of \textit{time
intervals} $\triangle t\equiv t-T_{1}$ at the end of which element $j$ fails.
As above, a value $\triangle t$ may be randomly chosen from this distribution
by use of the relation%
\begin{equation}
\triangle t=\tau_{j}^{\prime}[-\ln(1-x)] \label{e7}%
\end{equation}
where $x\in\lbrack0,1)$. Thus the next element to fail is that producing the
smallest member of the set $\{\triangle t_{j}\}$.

An alternative approach to simulating the system evolution is to
\textit{randomly choose} the next element to fail according to the set of
probabilities $\left\{  \tau_{k}^{-1}/%
{\textstyle\sum\nolimits_{j}}
\left(  \tau_{j}^{-1}\right)  \right\}  $, where element $k$ is one of the set
$\{j\}$, $\tau_{k}^{-1}/%
{\textstyle\sum\nolimits_{j}}
\left(  \tau_{j}^{-1}\right)  $ is the probability that element $k$ is the
next to fail, and the sum is over all $N$ currently surviving elements $j$
(remember that the values $\{\tau_{j}\}$ may change after every failure). This
failure occurs at the end of the time increment $\triangle t$, that may be
taken to be the \textit{average}\ value of the \textit{smallest} member of the
set $\{\triangle t_{j}\}$ (were the set to be calculated innumerable times),%
\begin{align}
\triangle t  &  =\left\langle \triangle t_{k}\frac{\tau_{k}^{-1}}{%
{\textstyle\sum\nolimits_{j}}
\tau_{j}^{-1}}\right\rangle =\frac{1}{N}\sum\limits_{k=1}^{N}\triangle
t_{k}\frac{\tau_{k}^{-1}}{%
{\textstyle\sum\nolimits_{j}}
\tau_{j}^{-1}}\nonumber\\
&  =\frac{1}{N}\frac{\sum\nolimits_{k}[-\ln(1-x_{k})]}{%
{\textstyle\sum\nolimits_{j}}
\tau_{j}^{-1}}=\frac{\left\langle -\ln(1-x_{k})\right\rangle }{%
{\textstyle\sum\nolimits_{j}}
\tau_{j}^{-1}}\nonumber\\
&  =\frac{1}{%
{\textstyle\sum\nolimits_{j}}
\tau_{j}^{-1}}\text{.} \label{e8}%
\end{align}

A different expression for $\triangle t$ may be found by noting that the
probability $p(\triangle t)dt$ that the next element to fail will do so during
the infinitesimal time interval $[\triangle t,\triangle t+dt]$ is%
\begin{equation}
p(\triangle t)dt=\sum\limits_{j=1}^{N}\left\{  p_{j}(\triangle t)dt\prod
\limits_{k(\neq j)=1}^{N}\left[  1-p_{k}(\triangle t)\right]  \right\}
\label{e9}%
\end{equation}
where the content of the curly brackets is the probability that element $j$
will fail during the infinitesimal time interval $[\triangle t,\triangle
t+dt]$, multiplied by the probability that no other element will fail during
$\triangle t$. Then%
\begin{align}
p(\triangle t)  &  =\sum\limits_{j=1}^{N}\left\{  \frac{1}{\tau_{j}}%
\exp\left[  -\frac{\triangle t}{\tau_{j}}\right]  \prod\limits_{k(\neq
j)=1}^{N}\exp\left[  -\frac{\triangle t}{\tau_{k}}\right]  \right\}
\nonumber\\
&  =\sum\limits_{j=1}^{N}\left\{  \frac{1}{\tau_{j}}\exp\left[  -\triangle t%
{\textstyle\sum\nolimits_{k=1}^{N}}
\tau_{k}^{-1}\right]  \right\} \nonumber\\
&  =\left(
{\textstyle\sum\limits_{j=1}^{N}}
\tau_{j}^{-1}\right)  \exp\left[  -\triangle t%
{\textstyle\sum\nolimits_{j=1}^{N}}
\tau_{j}^{-1}\right]  \text{.} \label{e10}%
\end{align}
As above, a value $\triangle t$ may be randomly chosen from this distribution
by use of the relation%
\begin{equation}
\triangle t=\frac{-\ln(1-x)}{%
{\textstyle\sum\nolimits_{j}}
\tau_{j}^{-1}} \label{e11}%
\end{equation}
where $x\in\lbrack0,1)$. The average value is given by Eq. (\ref{e8}), as expected.

The evolution of the system of elements is thus accomplished by selecting an
element to fail, and incrementing time accordingly. With each failure the
system enters a new state; obviously it cannot return to any old states. For
this reason it has been convenient to use the set of \textquotedblleft
lifetimes\textquotedblright\ $\{\tau_{j}\}$ corresponding to the surviving
elements $j$, rather than the set of transition rates $\{k_{i\rightarrow j}\}$
for transitions from the current state $i$ to the \textit{accessible} states
$j$ (the system will enter state $j$ if element $j$ is the next to fail). The
connection between the sets is made by noting that $%
{\textstyle\sum\nolimits_{j}}
\tau_{j}^{-1}=%
{\textstyle\sum\nolimits_{j}}
k_{i\rightarrow j}$, where the system is currently in state $i$ in either case.

By making this replacement in the denominators of Eqs. (\ref{e8}) and
(\ref{e11}), those equations for $\triangle t$ may be used for
\textit{equilibrium} systems (where the set of states and transition rates
between states don't change) as well. More generally, $\tau_{j}$ in the
equations above may be replaced by $k_{i\rightarrow j}^{-1}$ when it is
understood that the system is currently in state $i$.

For computational efficiency, it is most typical for a simulation to select
the transition from current state $i$ according to the set of probabilities
$\left\{  k_{i\rightarrow j^{\prime}}/%
{\textstyle\sum\nolimits_{j}}
k_{i\rightarrow j}\right\}  $ and calculate the transition time interval
$\triangle t$ using either Eq. (\ref{e8}) or (\ref{e11}), rather than to
calculate the set $\{\triangle t_{j}\}$ from the relation $\triangle
t_{j}=k_{i\rightarrow j}^{-1}[-\ln(1-x)]$ and choose the destination state
$j^{\prime}$ from the smallest of those values. When the \textit{distribution}
of simulation completion times is desired (for example, the time at which all
elements have failed), it is necessary to perform a large number of nominally
identical simulations, using Eq. (\ref{e11}) rather than Eq. (\ref{e8}) for
every transition time interval $\triangle t$. (Note that this distribution
will converge to a Gaussian distribution, according to the central limit theorem.)

More generally, evolution of a dynamic system is \textit{not} described by a
set $\{k_{i\rightarrow j}\}$ of time-independent rate constants. In this case
there is no set of probabilities $\left\{  k_{i\rightarrow j^{\prime}}/%
{\textstyle\sum\nolimits_{j}}
k_{i\rightarrow j}\right\}  $ from which to select the destination state
$j^{\prime}$. However, a set $\{t_{j}\}$ of transition times can still be
calculated from a PDF, giving the \textit{sequence} of events from an initial
time. For example, consider that the distribution of failure times for element
$j$ is given by the two-parameter ($\gamma$ and $\tau$) Weibull distribution%
\begin{equation}
p(t)=\gamma\frac{t^{\gamma-1}}{\tau^{\gamma}}\exp\left[  -\left(  \frac
{t}{\tau}\right)  ^{\gamma}\right]  \label{e12}%
\end{equation}
rather than by Eq. (\ref{e3}). For $\gamma>1$ this PDF decays faster than
exponentially and so might be appropriate for an element that
\textquotedblleft ages\textquotedblright, or accumulates damage over time.
(Indeed, the probability $\pi(t)dt$ that the element, having survived to time
$t$, will immediately fail is $\gamma(t^{\gamma-1}/\tau^{\gamma})dt$, meaning
that the probability of immediate failure \textit{increases} with time.) The
first and second moments of this distribution are, respectively,%
\begin{equation}
\left\langle t\right\rangle =\tau\frac{1}{\gamma}\Gamma\left(  \frac{1}%
{\gamma}\right)
\end{equation}
and%
\begin{equation}
\left\langle t^{2}\right\rangle =\tau^{2}\frac{2}{\gamma}\Gamma\left(
\frac{2}{\gamma}\right)
\end{equation}
where $\Gamma$ is the Gamma function. The counterpart to Eq. (\ref{e4}) is%
\begin{equation}
x(t)=\int\limits_{0}^{t}p(t^{\prime})dt^{\prime}=1-\exp\left[  -\left(
\frac{t}{\tau}\right)  ^{\gamma}\right]  \text{,} \label{e13}%
\end{equation}
so the failure time for element $j$ is%
\begin{equation}
t_{j}=\tau_{j}[-\ln(1-x)]^{1/\gamma_{j}} \label{e14}%
\end{equation}
where $x\in\lbrack0,1)$.

If the parameter values $\{\gamma_{j},\tau_{j}\}$ change with each element
failure (reflecting, say, more rapid aging), the PDFs $\{p_{j}(t)\}$ given by
Eq. (\ref{e12}) change accordingly. To calculate a new set $\{t_{j}\}$ of
transition times following an element failure at time $T$, we note that $x$
and $t$ must be related by%
\begin{equation}
\int\limits_{x(T)}^{x}dx^{\prime}=\int\limits_{T}^{t}p(t^{\prime})dt^{\prime}%
\end{equation}
where the lower limit $x(T)=1-\exp[-(T/\tau^{\prime})^{\gamma^{\prime}}]$ is
obtained from Eq. (\ref{e13}), and $\gamma^{\prime}$ and $\tau^{\prime}$ are
the parameter values for $t>T$. Then the counterpart to Eq. (\ref{e4}) is%
\begin{align}
x(t)  &  =x(T)+\exp\left[  -\left(  \frac{T}{\tau^{\prime}}\right)
^{\gamma^{\prime}}\right]  -\exp\left[  -\left(  \frac{t}{\tau^{\prime}%
}\right)  ^{\gamma^{\prime}}\right] \nonumber\\
&  =1-\exp\left[  -\left(  \frac{t}{\tau^{\prime}}\right)  ^{\gamma^{\prime}%
}\right]
\end{align}
so the failure time for element $j$ is%
\begin{equation}
t_{j}=\tau_{j}^{\prime}[-\ln(1-x)]^{1/\gamma_{j}^{\prime}}%
\end{equation}
where $x\in\lbrack x(T),1)$.

\section{Discussion}

In physics and materials research, the kMC method is most often used for
simulation of non-deterministic, thermally activated processes, as these give
time-independent rate constants (time-independent probabilities) for
transitions between states. A similar, but more complex, example to the adatom
diffusion calculation mentioned in the Introduction is diffusion of vacancies
and interstitial atoms in a grain boundary \cite{r1}, where the transitions
between states may occur by very complex mechanisms (for example, by concerted
motion of atoms). However the rates $k$ are of the simple form%
\begin{equation}
k_{i\rightarrow j}=\nu\exp\left[  -\frac{E_{m}}{k_{\text{B}}T}\right]
\end{equation}
where $E_{m}$ is the energy barrier between the initial state $i$ and the
final state $j$ (which the system may overcome with thermal energy supplied by
local temperature fluctuations), $k_{\text{B}}$ is the Boltzmann constant, $T$
is the average temperature of the system, and $\nu$ is the frequency with
which the system attempts to make the transition. The latter quantity gives
the time scale for the microscopic process, that is needed for the kMC simulation.

Besides its natural application to atomic-scale processes, the kMC method has
been used to simulate the time-dependent damage and failure of material under
stress. For example, Curtin \textit{et al.} \cite{r2} considered a spring
network model where the rate $r$ of failure at site $j$ is%
\begin{equation}
r_{j}(t)=A\sigma_{j}(t)^{\eta}%
\end{equation}
where $\sigma_{j}(t)$ is the local stress at time $t$, and the exponent
$\eta>1$ ensures a nonlinear relationship between damage rate and stress as is
the case for creep. The time between failures was calculated by Eq.
(\ref{e8}), after which all the local stresses $\{\sigma_{j}\}$ and failure
rates $\{r_{j}\}$ were recalculated. In contrast, Harlow \textit{et al.}
\cite{r3} considered a polycrystal under stress, and randomly chose, from a
probability distribution that included the local stress as a parameter, a
time-to-failure $t_{j}$ for each grain boundary facet $j$. Then time was
advanced by the amount of the smallest member of the set $\{t_{j}\}$, the load
supported by that failed facet was redistributed to adjacent facets, and a new
set $\{t_{j}\}$ was obtained from new distributions. In a similar way Andersen
and Sornette \cite{r4} considered a fiber rupture model where failure of each
fiber affected the failure-time probability distributions for the remaining fibers.

Effective use of the kMC method obviously requires that all important states
of the system, and all important transitions between them, be identified. How
to ensure this, and other technical and computational (i.e., practical) issues
are addressed in a review by Voter \cite{r5}. Another problem is presented by
\textquotedblleft basins\textquotedblright\ of states, which are subsets of
mutually accessible states from which escape is a very rare event. This of
course defeats the purpose of the kMC method. However Van Siclen \cite{r6} has
recently shown how to calculate the residence time for a system trapped in a
basin, under the assumption that the system has equilibrated in the basin
(which is to say, there is no correlation between the entry and exit states).

To conclude, the kinetic Monte Carlo method is a powerful technique for
simulating the dynamics or evolution of non-deterministic systems. Thus it
should have applications beyond the physical sciences, for example to biology,
ecology, risk assessment, and finance.

\begin{acknowledgments}
This work was supported in part by the INL Laboratory Directed Research and
Development Program under DOE Idaho Operations Office Contract DE-AC07-05ID14517.
\end{acknowledgments}

\end{document}